\documentclass[10pt,twocolumn]{revtex4}
\usepackage{graphicx,color,comment,amsmath,mathtools}
\usepackage{amsfonts}

\sffamily

\begin{document}

\thispagestyle{empty}

\title{Towards Inverse Modeling of Intratumor Heterogeneity}

\author{B. Brutovsky}
\affiliation{Department of Biophysics, Faculty of Science, Jesenna 5, P.~J.~Safarik University, 04154 Kosice, Slovak Republic}

\author{D. Horvath}
\affiliation{Centre of Interdisciplinary Biosciences, Faculty of Science, Jesenna 5, P.~J.~Safarik University, 04154 Kosice, Slovak Republic}

\begin{abstract}
Development of resistance limits efficiency of present anticancer
therapies and preventing it remains big challenge in cancer research.
It is accepted, at intuitive level, that the resistance emerges
as a consequence of cancer cells heterogeneity at molecular, genetic
and cellular levels. Produced by many sources, tumor
heterogeneity is extremely complex time dependent statistical
characteristics which may be quantified by the measures defined in many
different ways, most of them coming from statistical mechanics.
In the paper we apply Markovian framework to relate population heterogeneity
with the statistics of environment.
As, from the evolutionary viewpoint, therapy corresponds to a purposeful
modification of the cells fitness landscape, we assume that understanding
general relation between spatiotemporal statistics of tumor microenvironment
and intratumor heterogeneity enables to conceive the therapy as the inverse
problem and solve it by optimization techniques.
To account for the inherent stochasticity of biological processes
at cellular scale, the generalized distance-based concept was applied
to express distances between probabilistically described cell states
and environmental conditions, respectively.
\end{abstract}

\maketitle

\section{Introduction}

Intratumor heterogeneity (ITH), referring to biological differences between
malignant cells within the same tumor, is considered to be a major obstacle
in successful eradicating tumors~\cite{Marusyk2013_Science}.
While normal cells respond very similarly to drugs, mechanisms of resistance
of cancer cells are extremely diverse \cite{Gottesman2002,Altschuler2010},
which poses real challenge for targeted therapies.
Therefore, the development of novel effective cancer treatment strategies
requires deep understanding causes and consequences of high variability
of cancer cells, and, eventually, its control.

ITH at the level of DNA sequences (below denoted as genetic)
is well understood as necessary prerequisite of cancer evolution.
On the other hand, emerging evidence supports the view, that the
ability of cancer cells to switch between alternative states
(or phenotypes) without the change of their genotype, known as
plasticity, may be essential in many cancer types \cite{Marjanovic2013}.
The role of this non-genetic (or epigenetic) part of ITH
in cancer progression is, however, from evolutionary viewpoint
less obvious \cite{Huang2013_CancerMetastasisRev}.
The evidence accumulates, that dynamic and reversible phenotype
plasticity may constitute an "escape route" for cancer cells
which may become more invasive and resistant to therapy \cite{Kemper2014}.

Cancer research usually concentrates on molecular details,
implicitly presuming predominance of determinism in cancer
causation. Taking into account that ITH results from specifically
altered biochemical interactions of the cells with their
environment \cite{Tam2013,ElShamy2013}, the effort to
understand specific biochemistry of cancer cell for its 
therapeutic application is understandable.
However, as ITH {\it by definition} represents collective
property of the cells population, its role is conceived with
difficulty from the single-cell viewpoint.
The recognition that the stochasticity of molecular processes itself
induces heterogeneity of responses to drugs, which may have clinical
impact even in the case of genetically identical cells under identical
physical conditions \cite{Saunders2012}, underlines necessity to integrate
stochastic aspect of cancer progression into cancer models. 

Being driven by the two components, genetic and environmental, development
of ITH becomes extremely complex phenomenon, which may be quantified by
different measures, most of them coming from statistical mechanics,
e. g. entropy concept \cite{Buckland2005,Keylock2005,Mendes2008}.
Transforming ITH into a tractable and computable property
of the population of cells provides a rigorous starting point for
developing mathematical cancer models and simulations \cite{Preziosi2003}.

In the paper we presume, that statistics of ITH plays in cancer
initiation and progression important role {\it per se} and,
moreover, it may be studied separately from underlying it biochemistry.
Instead of trying to be (too) detailed in some of the aspects,
we put the emphasis on the integration of the universal 
evolutionary features into the overall scenario.
Applied Markovian-based framework enables to study universal
causative role of environmental dynamics on the heterogeneity
of the population of asexually reproducing units ("cells").
Cell states heterogeneity is bound with environmental statistics by
making transition probabilities dependent on generalized distances
between probability distribution functions corresponding to environment
and the cell states, respectively.

\section{Master equation approach to cell states heterogeneity}

It is often reported that tumor propagating cells are maintained by stochastic,
rather than deterministic, mechanisms which are, at least, partially reversible
\cite{Maenhaut2010,Chang2008,Quintana2010,Sharma2010,Hoek2010,Chaffer2011}.
It was observed \cite{Gupta2011} that the population of human breast
cancer cells consists of three phenotypically different sub-populations
(consisting of stem, basal and luminal cells, respectively).
By studying dynamics of these cell type fractions it was found, that they
stay, under stationary conditions, in equilibrium proportions~\cite{Gupta2011}.
Moreover, if the cancer cells population was purified for any of the three cell
types, the equilibrium was re-established too rapidly to be explained by
differential growth rates of the respective cell types fractions
and it was proposed that phenotypic equilibrium was maintained
by stochastic transitions between different cell states. Assuming that
the transition rates per unit time are, under fixed genetic
and environmental conditions, constant, the cell transitions dynamics
was identified with Markovian process~\cite{Gupta2011}.
Strong motivation for this comes from physics, where Markovian processes
are routinely applied to model dynamical systems which are, at any given
time, exactly in one from discrete number of states $\{1,2,\ldots, N\}$,
and where the transitions between states are treated probabilistically.   
Within Markovian formalism, the continuous time $t$ variation of the probability
obeys well known first-order phenomenological master equation
\begin{equation}
\frac{{\rm d} p_i}{{\rm d}t}=  \sum_{k=1}^N W_{k i} \, p_k -  \sum_{k=1}^N W_{i k} \, p_i\,,
\label{Master_Equation}
\end{equation}
where $p_i(t)$, $ i=1,2,\ldots, N,$ are probabilities that the system is
in the $i$-th state and $W_{k i}$, $ k= 1, 2, \ldots, N$ are transition probabilities 
$k$-th to the $i$-th state per unit time. The underlying principle of the
above equation, stating that the appropriate constant transition probabilities 
may produce physically correct stationary distributions, has been exploited
in the design of Monte Carlo importance sampling simulation techniques
\cite{Binder1988}.

Identification of the cell-state dynamics with Markovian process
enables to study statistical aspects of population dynamics separately
from the details of underlying it biochemistry, which stay hidden
in the probabilities of transitions between states.
Despite the fact, that biochemical
processes behind the respective transitions are very probably interdependent,
huge complexity of the problem leaves the opportunity to get, in principle,
any equilibrium distribution of the cell states by many alternative transition matrices.
Consistently with this, many paths and mechanisms of transitions between cell states
('phenotype switching') are observed at molecular, genetic and expression
levels~\cite{Choi2008,Raj2008,Eldar2010,Liberman2011,Pujadas2012} and theoretically
studied~\cite{Kussell2005,Kussell2005a,Acar2008,Frankenhuis2011,Libby2011,Fudenberg2012,Fedotov2007,Fedotov2008}.

It is well accepted that each cancer case represents
evolutionary process progressing during the individuum's lifetime
\cite{Nowell1976,Merlo2006,Greaves2007}.
A range of studies suggests that phenotype heterogeneity
results from the evolutionary pressure to keep gene expression in tune with
physiological needs dictated by the environment \cite{LopezMaury2008}.
Below we construct Markovian-based formal framework to integrate
phenotype heterogeneity into evolutionary scenario using the above model
by Gupta et al.~\cite{Gupta2011} as starting point.
Within the framework, phenotype heterogeneity is naturally identified
with the limiting distribution of states of the respective Markovian
process. Consequently, as the limiting distribution unambiguously results
from the transition probabilities (summarily denoted as the transition
matrix), the evolutionary pressure imprints them into the genes.
Apart from being hard-wired in the genes, the transition probabilities 
may be influenced by instantaneous microenvironment as well.
Regarding the environment sensing, the transition (or switching)
is usually termed as 'responsive' if it occurs as a direct
response to some environmental stimulus, or 'stochastic' if no direct
environment stimulus is present \cite{Kussell2005}.
Being demonstrated that population of breast cancer cells purified
for one of the stable cell types converges in stationary conditions
gradually to original (equilibrium) phenotypic fractions~\cite{Gupta2011}
(instead of immediate leap to phenotypically homogeneous population)
indicates that responsive switching is not the exclusive cause
of the transitions between the states.

The causation of transitions crucially predetermines mathematical
form of the transition probabilities and limiting distribution.
When $W_{k i}$ are constant (due to the stationarity of environment
or the cell's insensitivity to environmental non-stationarity), 
the cell states dynamics represents Markovian process.
When the influence from the environmental non-stationarity
cannot be neglected, $W_{k i}(t)$ are time-dependent and the process
becomes non-stationary (time-inhomogeneous Markov process).
In the next we assume that evolving population is influenced by time-varying
environment and we focus on the inherent structure of time-dependent
$W_{k i}(t)$ at phenomenological level. Below proposed structure of
$W_{k i}(t)$ preserves Markovian property of the cell
state dynamics and, on the other hand, enables
to study non-equilibrium phenomena which reflect temporal variability
of the environment.

The probability of transition between two states typically derives from
the distinction in some of their characteristics, such as energy levels
in Metropolis Monte Carlo method in statistical physics \cite{Binder1988}.
Non-stationarity of transition probabilities due to environment
fluctuations prevents the system from reaching limiting distribution
consisting of a few unique states.
However, when the fluctuations of environment are not correlated with
the environmental average, one can intuitively replace the concept of
limiting distribution with the notion of probability
distribution represented by the superposition of appropriately approximated
'peaks' of nonzero width around the lines corresponding to the 'pure' states
which would result from the stationary (evolutionary tuned) transition matrix.
Consequently, the question arises what is the quantity the probability
of which is distributed. Regarding the main aim of this study, which
is proposing formal framework enabling to explore how environment
statistics exerts evolutionary pressure on the statistics of evolving
population, the probability distribution relates to the effective
parameter integrating all the relevant environmental factors which
influence transition probabilities, below referred to as 'environmental cue'.
It relates to the environment itself, which is viewed as its donor, as well
as to the cell states, which plays the role of its eventual recipient.
More formally, two probability distributions for environmental cue may
be constructed, the former related to the environment, the latter
related to the cell states, respectively.

The next question is, how to measure the distinction between the states
that are described only probabilistically. For that, we apply the term
'distance' in its broad mathematical meaning, as a distance between
two probability distributions. In the next, we express
the transition matrix in the terms of generalized distances between
the probability density functions corresponding
to the environment and to the respective state, and between
the probability density functions corresponding to the two respective
cell states.
Here proposed formalism is consistent with the dynamical
system conceptualization \cite{Zhou2012}, where the cell states
were epitomized by the respective attractors distributed around
stable states in epigenetic landscape (see section III).

Biological relevance of the formalism results from the flow
cytometry experiments, where the phenotypic distributions
of cell populations are the typical outputs \cite{Altschuler2010,Chang2008}.
The distributions are not the artefacts caused by the imperfection
of experimental procedures, but they reflect phenotypic gene
expression noise \cite{Elowitz2002}, which is intensively studied
authentic biological phenomenon\cite{Eldar2010}.

The concept of attractor is, however, not context-free. 
Therefore, to continue in this conceptualization, we provide
more clarity about what we mean when talking about attractors
of deterministic systems accepting some degree of indeterminism.
Obviously, deterministic and stochastic systems have different
properties, and should be treated separately.
So far we have continued without mentioning problem with the stochastic
attractor framework originally conceived as deterministic,
although attractor is also pertinent to stochastic systems.
Our approach avoids purely mathematical concept of pullback attractor 
and process~\cite{Crauel97} which are actually of little relevance for our work. 
Instead, we prefer more intuitive picture where small noise
perturbations induces random switching between (stable) coexisting point 
attractors of different relative depth. Such scenarios are also conceivable 
in computational neuroscience in modeling multistable perception~\cite{Braun2010}. 
The transitions between attractors can be best characterized 
by the transition probabilities~\cite{Stiller1992}. Our formalism 
is to substantial extent influenced by the study of developmental
transitions~\cite{Zhou2012}, where dynamical features of attractors 
are comprised in the quasi-potentials of specific depth, 
while the transition probabilities 
between attractors are defined in analogous way as thermally activated 
transitions between equilibrium states. 

In the next, we apply the above considerations to construct
phenomenological relation between transition probability and
the match of environment and population statistics, both
expressed by the probability distribution of the above
environmental cue.
Our phenomenological model postulates, that transitions are associated with
matching conditions of the attractor distributions comprised in   
\begin{equation}
W_{i j} = C \exp\left( \alpha (\, f_j - f_i )) \, \exp(-\lambda\,  d^2_{i j} \right)\,,
\label{squared_dist}
\end{equation}
where $f_i$ expresses the measure of attractivity of the $i$-th attractor
under instant environmental factors; its sign will be discussed
later within the relevant biological context. The amplitude parameter
$\alpha$ determines relative strength of this environmental influence.
The second term, $\exp(-\lambda\, d^2_{ij})$, manifests dependence
of the transition probability on the generalized distance, $d_{ij}$, between
the attractors $i$ and $j$. The distance appears in (\ref{squared_dist})
in squared form in analogy with the transition term for diffusion of random walk process.
In this context $\lambda$ parameter represents reciprocal value of
the diffusion coefficient.

The constant $C$ simply stems from the normalization condition 
\begin{equation}
\sum_{j=1}^N W_{i j}=\frac{1}{\tau}\,,
\end{equation}
where $\tau$ is the specific time scale of the transitions. 
The parameter is assumed to be much smaller than the evolutionary
time scale.  
Then the normalized form of $W_{i j}$ may be written as  
\begin{equation} 
W_{i j} =  \frac{1}{\tau} \, \frac{ \exp(\alpha f_j - \lambda d^2_{i j}) }{\sum_{s=1}^N 
\exp(\alpha f_s - \lambda d^2_{i s}) }\,. 
\end{equation}
Suppose, that the probability density function of the environmental cue is
parametrized by the single parameter or parameters comprised in $\theta^{\rm e}$.
Similarly, the probability density function of the environmental cue
associated with the $i$-th attractor is parametrized by $\theta_i$,
$i=1,2,\ldots, N$. The impact of the $i$-th attractor is proportional
to the generalized distance of $\theta_i$ from the current probability
distribution which reflects environmental conditions expressed
by $\theta^{\rm e}$. To sum up, the effect of environment may be
quantified by a squared generalized distance $d^2(\theta_i,
\theta^{\rm e}(t))$, normalized, without the loss of generality,
to interval ranging from $0$ to $1$. This tendency is captured
by the phenomenological equation
\begin{equation} 
f_i(t) =  f_{\rm A} \, \left(\, 1 - d^2(\theta_i, \theta^{\rm e}(t))\, \right)\,, 
\end{equation} 
where $f_{\rm A}>0$ is the amplitude common to all attractors,
the sign of which follows from the biological
context. In evolutionary biology, populations of isogenic
individuals evolving in time-varying environments typically
develop bet-hedging strategy, which means that the statistics
of states is coupled with the statistics of environment
\cite{Donaldson2008}. To reflect biological relevance, i. e. that
more accurate matching of the state statistics with the statistics
of environment represents comparative advantage, and,
at the same time, to stay consistent with the Eq.~(\ref{squared_dist}),
we postulated $f_{\rm A}>0$.
Within the above biological context, $\theta_i$ is assumed
to be fixed (being already evolved), while the parameters
of the environment statistics comprised in
$\theta^{\rm e}(t)$ are allowed to vary. After the substitutions,
$W_{i j}$ can be simply rewritten into the following form
\begin{equation} 
\begin{split}
W_{i j}(t, 
\{\theta_i\}_{i=1}^N,   
\theta^{\rm e}(t)) =
\ \ \ \ \ \ \ \ \ \ \ 
\\ 
\\
\frac{1}{\tau} \frac{ \exp\left[-  \beta d^2({\theta_j, \theta^{\rm e}}(t)) - 
\lambda d^2({\theta_i,\theta_j}) \right]}{\sum\limits_{s=1}^N  \exp 
\left[ -  \beta d^2({\theta_s, \theta^{\rm e}}(t)) -  \lambda d^2({\theta_i,\theta_s})
\right]}\,,
\end{split}
\label{sensing}
\end{equation}
where new parameter $\beta$ replaces the product $\alpha f_{\rm A}$.

Within context of the above classification, the parameters comprised 
in Eq.~\ref{sensing} may be interpreted as follows. The parameter
$\beta$ expresses dependence of the transition probabilities on
instant environment and corresponds to responsive switching,
and $\lambda$ is related to the probability of stochastic
switching. The constants $\beta$, $\lambda$ stem purely
from genetic basis which was fixed during long evolutionary history.

\section{Generalized distance between attractors}

Accordingly to the instructive conceptualization~\cite{Huang2013_CancerMetastasisRev},
each point in the genomic landscape (i.~e.~genome) provides epigenetic landscape
of unique topology, which, due to its mathematical complexity, contains many
stabilizing areas of space (attractors) around stable (or equilibrium) states.
Transitions between attractors dominate in complex system's behavior 
at its relevant time scales and represent additional force to the component
of the force which follows gradient in the (quasi-potential) epigenetic
landscape~\cite{Zhou2012}.
The system may contain countable set of attractors of different types
adjoining each other and, intuitively, the probability of transitions between
specific states depends on the depth and form of the respective 
attractors. Therefore, to put forward the above outlined conceptualization,
the distributions (which are attractors in the functional space) must be specified.
The arguments given here may be applied to simple, as well as highly
complex parametrization.

Below we presume the attractors with normally distributed fluctuations.
In such case we assume $\theta_i  \equiv (\mu_i,  \sigma_i)$, $i=1,2,\ldots, N$,
where $\mu_i$  denotes the mean  of the selected factor and $\sigma_i$
is its dispersion. Analogously, for the environment  we assume the
parametrization $\theta^{\rm e}(t) \equiv (\mu^{\rm e}(t),\, \sigma^{\rm e}(t))$.
Dissimilarity between the pairs of normal distributions is characterized
by the {\em Hellinger distance}~\cite{Hellinger1909}. 
This original forms are modified by the regularization (see Appendix).
In agreement with the assumptions and parametrization of the model
discussed, we use regularized Hellinger distances in two contexts:

i) the inter-attractor form
\begin{equation} 
\label{dist1}
\begin{split}
d^2_{i j} \equiv d^2(\mu_i, \sigma_i, \mu_j, \sigma_j,\epsilon) = \ \ \ \ \ \ \ \ \ \ 
\ \ \ \ \ \ \ \ \ \ \ \ \ \ \ \ \ \ \ \ \ \ \ \ \\
1 - \sqrt{\frac{2 (\sigma_i+\epsilon) (\sigma_j+ \epsilon)}{ 
(\sigma_i+\epsilon)^2 + (\sigma_j+\epsilon)^2}}
\ \ \ \ \ \ \ \ \ \ \ \ \ \ \ \ \\
\times  \exp \left(- \frac{1}{4} \frac{ (\mu_i - \mu_j)^2 
}{  (\sigma_i+\epsilon)^2 +  (\sigma_j +\epsilon)^2 }\right)  
\end{split}
\end{equation} 

and, ii) attractor-environment form
\begin{equation}
\label{dist2}
\begin{split}
d^2_i \equiv d^2(\mu_i,\sigma_i,\mu^{\rm e}(t), 
\sigma^{\rm e}(t),\epsilon) = \ \ \ \ \ \
\ \ \ \ \ \ \ \ \ \ \ \ \ \ \ \ \ \ \ \ \ \ \ \ \ \ \ \ \\
1 -  \sqrt{\frac{2  (\sigma_i +\epsilon) (\sigma^{\rm e}(t)+\epsilon)}{ 
(\sigma_i + \epsilon)^2 + (\sigma^{\rm e}(t) + \epsilon)^2}}
 \ \ \ \ \ \ \ \ \ \ \ \ \ \ \ \ \ \ \ \ \ \\
\times \exp \left( 
- \frac{1}{4} \frac{ (\mu_i-\mu^{\rm e}(t))^2 }{ 
(\sigma_i+\epsilon)^2 +  (\sigma^{\rm e}(t) + \epsilon)^2} \right)\,. \ \ \
\end{split}
\end{equation} 
All the distances are regularized by the unique additive parameter
$\epsilon>0$, which plays the role of additional contribution to dispersion or
determines respective generalized geometrical context.
The functions are homogeneous 
of the order zero in the following sense 
\begin{equation}
\begin{split}
d^2( \xi \mu_i,  \xi  \sigma_i, \xi \sigma_j, \xi  \sigma_j, \xi  \epsilon) =
d^2(\mu_i ,\sigma_i, \mu_j, \sigma_j, \epsilon)\,,
\\
d^2(\xi \mu_i,\xi \sigma_i, \xi \mu^{\rm e}(t), \xi \sigma^{\rm e}(t), 
\xi \epsilon) 
=
\ \ \ \ \ \ \ \ \ \ \ \ \ \ \ \ \ \ \ \ \ \ \\
d^2(\mu_i ,\sigma_i, \mu^{\rm e}(t), \sigma^{\rm e}(t),\epsilon) \,. 
\end{split}
\end{equation}
The above equation trivially induces scale invariance of the transition matrix.
The independence on the scaling parameter $\xi$ implies generality of the
conclusions derived from the particular calculation, which makes relevant
proportions of the parameters $\mu_i$ and $\sigma_i$, and time dependencies
$\mu^{\rm e}(t)$ and $\sigma^{\rm e}(t)$ instead of their values themselves.

The idea of regularization is to keep dependence upon the
$\mu_i$, $\mu_j$ and $\mu^{\rm e}(t)$ even in the
anomalous situation when the dispersions vanish
\begin{eqnarray}
d^2(\mu_i,0,\mu_j,0,\epsilon) &=& 1 - 
\exp\left(-\frac{(\mu_i-\mu_j)^2}{ 
8 \epsilon^2}  \right)\,, 
\\
d^2(\mu_i,0,\mu^{\rm e},0,\epsilon) &=&  
1 - \exp\left( -  \frac{(\mu_i-\mu^{\rm e}(t))^2}{ 8 \epsilon^2} \right)\,. 
\nonumber
\end{eqnarray}
The above relationship indicates, that only $\epsilon>0$ guarantees sensitivity of 
the distance to the mean values also 
in the case of zero dispersions $\sigma_i$
and $\sigma_j$.

Now, instead of interest in particular attractor, we continue
with the construction of the probabilistic model for the response
probability density function $P(x,t)$, representing the system
of attractors, along some cumulative cell state characteristics $x$, 
playing formally the role of interpolation variable. 
Note that $x$ is assumed to have the same origin (i. e. the same meaning, dimension
and unit) as $\mu_i$ and $\sigma_i$. 
Following above conceptualization,
we express $ P(x,t) $ as a probabilistic multimodal Gaussian mixture model 
\begin{equation}
P(x,t)=\sum_{i=1}^N p_i(t) {\phi}_{\rm g}( x ; \mu_i, \sigma_i) 
\label{gaussmix}
\end{equation}
based on the convex combination of $N$ Gaussian response probability density functions 
\begin{eqnarray}
{\phi}_{\rm g}(x ;  \mu,  \sigma) = \frac{1}{\sqrt{2 \pi} \sigma} \exp \left( -\frac{ (x-\mu)^2}{ 
2 \sigma^2}\,\right)\,.
\label{gaussdens}
\end{eqnarray}
In this formula, the previously introduced probabilities $\{\,p_i(t)\}_{i=1}^N$
play the role of mixture weights. At the level of description using $P(x,t)$,
passing along the only parameter $x$ comprises all the key observable
statistical characteristics of the system of attractors, whereas
$\{\,\,\mu_i, \sigma_i\}_{i=1}^N$ pairs may be viewed as partially hidden.

The reader interested in consistency with the classical view of evolutionary biology may find
it interesting, that instant fitness of the genetically identical cell
population at given conditions $x$ may be constructed as a monotonous
function of $\ln P(x,t)$ argument. 
The normalization 
\begin{equation}
\int_{-\infty}^{\infty} P(x,t) {\rm d}x  =  \sum_{i=1}^N p_i(t) \, 
\int_{-\infty}^{\infty}\,{\rm d}x \, \phi_{\rm g}( x; \mu_i, \sigma_i) = 1 
\end{equation}
stems from the obvious relations  
\begin{equation}
\int_{-\infty}^{\infty} \phi_{\rm g}(x; \mu_i, \sigma_i) = 1\,, \qquad \sum_{i=1}^N p_i(t) = 1 \,. 
\end{equation}
This model enables to determine the total mean, $\mu^{(1)}(t)$, and the dispersion, $\sigma(t)$, 
of $P(x,t)$ as follows
\begin{eqnarray}
\mu^{(1)}(t) &=& \sum_{i=1}^N p_i(t) \mu_i\,,
\\
\sigma^2(t) & =&  \sum_{i=1}^N p_i(t) \left[  
(\mu_i - \mu^{(1)}(t) )^2 
+ \sigma_i^2 \right]\,.
\end{eqnarray}
The latter characteristics, $\sigma(t)$, describes heterogeneity in attractors
occupancies. We note that the assumption of gaussianity is not an ultimate
requirement for the eventual applicability and functionality of the method. 
Any single peak function that resembles Gaussian, such as Lorentzian
or generalized exponential distributions~\cite{Hesse2006}, may be used
to specify the behavior in the vicinity of the fixed point attractor
to achieve faster convergence or more accurate mixing.

We note that one cannot {\it a priori} exclude
other shapes for characterizing the complex attractors.
Despite being chosen mainly because of its simplicity, gaussianity
of $\phi_{\rm g}$ is not the limiting aspect of this study
owing to the remarkable potential of the probability mixture models
to display higher-order moments
\begin{equation} 
\Delta\mu^{(k)}(t) = 
\sum_{i=1}^N p_i (t)\,\phi_{\rm g}^{(k,i)}(t)\,,
\,\,\, k=1, 2, \ldots 
\end{equation}
with variable coefficients 
\begin{equation}
\phi_{\rm g}^{(k,i)}(t) \equiv 
\int_{-\infty}^{\infty} 
{\rm d}x \,(x-\mu^{(1)}(t)\,)^k
\,\phi_{\rm g}(x; \mu_i,\sigma_i)
\end{equation}
related to the skewness $\Delta\mu^{(3)}(t)/\sigma^{3/2}(t)$ 
and kurtosis $\Delta\mu^{(4)}(t)/\sigma^4 (t)$ shape-related measures.
Consequently, even fat tails can arise as the side effect of statistical
mixing of the distributions. Similarly, multimodality
of many characteristics of complex biological systems is often observed
\cite{Hull1989}.
Note that the asymptotic approaching to a multimodal distribution 
is obvious in Fig.~\ref{fig:init} as well.
 
Moreover, alternative and, from the information theoretic
perspective, more conventional measures of the system of attractors
may be used.
If specificity of attractors  (comprised in 
$\phi_{\rm g}$) is not taken into account one may use the 
classical Shannon's  definition
\begin{equation}
S_{\rm sh}(t) = -\sum_{i=1}^N p_i(t) \ln p_i(t)\,. 
\label{S_sh}
\end{equation}
On the other hand, if one focuses on the stochastic transitions between
attractors, then more appropriate measure is Markovian chain
rates entropy
\begin{equation}
S_{\rm mr}(t) = - \sum_{i,j}^{N\times N} p_i(t) W_{i j}(t) \ln W_{i j}(t)\,. 
\label{S_mr}
\end{equation}
Due to complexity of the presented model, temporal behavior of both
the entropy measures can only be provided by numerical integration.
In the near future it would be also interesting to investigate Markovian
framework of phenotypic switching regarding non-ergodicity related
to the occurrence, non-occurrence or blocking of specific attractors.

Despite purely conceptual essence of the presented model, integrating
distinguished cancer-relevant features, such as increased
heterogeneity, phenotype switching and cell-to-cell
variability in the mathematical framework, it can eventually be
applied to analyze experimental data as well. We presume that once
the distributions along an appropriately chosen environmental cue
and the frequencies of transitions between the respective attractors
are known, the parameters $\beta$ and $\lambda$ (Eq.~\ref{sensing})
may be inferred, indicating relative contributions of responsive
and stochastic switching, respectively.
As a starting point, one could analyze two-state systems, which are
intensively studied at experimental \cite{Solopova2014} and theoretical
\cite{Kuwahara2012} levels.

\section{Numerical illustration of the system behavior} 

\begin{figure}[h!]
\includegraphics[scale = 0.8]{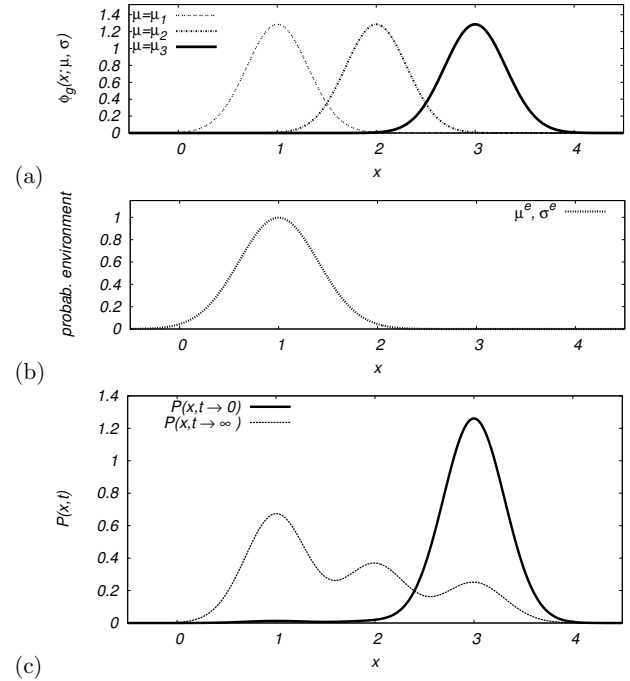}
\caption{The $N=3$ state system and the non-equilibrium dynamic behavior of 
the probability density function $P(x,t)$. 
Part~(a) The Gaussian functions $\phi_{\rm g}(x; \mu_i, \sigma_i)$, $i=1,2,3$. 
Part~(b) The Gaussian function corresponding to the environment with statistical properties 
defined by the parameters $\mu^{\rm e}=1$ and $\sigma^{e}=0.4$. 
Part~(c) The dynamics represented by $P(x,t)$ obtained using Eq.(\ref{gaussmix}).} 
\label{fig:init}
\end{figure}

\begin{figure}[h!]
\includegraphics[scale = 0.65]{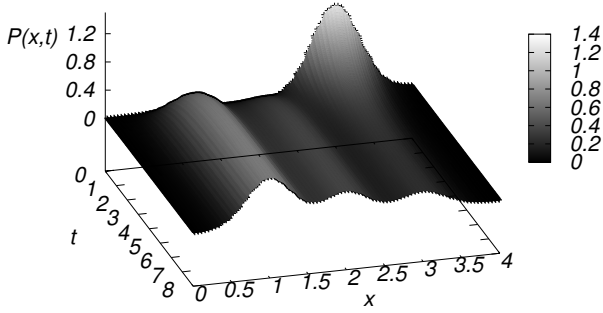}
\caption{The supplementary 3d view showing the variation of $P(x,t)$ until the limiting distribution is attained.
Calculated for the same conditions as in Fig.~\ref{fig:init}. 
The formation of "tree-hill" (long-term) asymptotics is visible.}  
\label{fig:3dP}
\end{figure}

To illustrate behavior of the above model, numerical simulation of its dynamics
was performed for the selected values of parameters. The model system is built
using Eqs.~(\ref{gaussmix},~\ref{gaussdens}) and the dynamics presribed
by the Eq.~(\ref{Master_Equation}) with the transition probabilities (Eq.~\ref{sensing})
and generalized distances (Eqs.~\ref{dist1},~\ref{dist2}) updated in due time.
The model system consists of $N=3$
states characterized by the function $\phi_g(x;\mu_i,\sigma_i)$, $i=1,2,3$
(Eq.~\ref{gaussdens}), each of them defined by the evenly spaced mean values
$\mu_1=1$, $\mu_2=2$, $\mu_3=3$ and identical dispersions
$\sigma_1 = \sigma_2 = \sigma_3 = 0.3$ (Fig.~\ref{fig:init}a).
The Gaussian function corresponding to the environment (Eq.~\ref{gaussdens}) 
was defined
by the parameters $\mu^{\rm e}=1$ and $\sigma^{e}=0.4$ (Fig.~\ref{fig:init}b).
To obtain $P(x,t)$ (Eq.~\ref{gaussmix}), numerical solutions $p_i(t)$
of Eqs.~(\ref{Master_Equation}) and (\ref{gaussmix}) were used.
The update of instantaneous matrix elements $W_{i k}(t)$ was 
performed using Eq.~(\ref{sensing}) with the parameters
$\beta=1$, $\lambda=0.1$, $\tau=1$ and $\epsilon=0.01$, and
substitutions $d^2({\theta_i, \theta^{\rm e}}) =
d^2(\mu_i,\sigma_i,\mu^{\rm e},\sigma^{\rm e})$ (Eq.~\ref{dist2})
and $d^2({\theta_i,\theta_j}) = d^2(\mu_i,\sigma_i,\mu_j,\sigma_j)$
(Eq.~\ref{dist1}).
The integration was performed by Euler method (integration step $\Delta t=0.01$)
under the normalization condition $\sum_{i=1}^{N} p_i = 1$.
The influence of stationary environmental statistics was studied
(see Fig.~\ref{fig:init}).
At initialization, the system is localized around the distribution
$P(x,t=0) \sim \phi_g(x;\mu_3,\sigma_3)$ which can be represented
by the initial condition of Eq.(\ref{Master_Equation}):
$p_1(0) = 0.01$, $p_2(0) = 0.01$, $p_3(0)=0.98$.
It means that the initial distribution $P(x,t=0)$ is very distinct from 
the environmental preference (given by $\mu^{\rm e}=1$ and $\sigma^{e}=0.4$).
In this nonequilibrium situation the model system is under environmental
"pressure" to increase the $p_1(t)$ fraction.
The expectation is confirmed by the numerical solution of the master
equation (Eq.~\ref{Master_Equation}) converging to the long-term values 
$p_1^{\ast} = 0.52 \gg p_1(0)$, $p_2^{\ast}=0.28$, $p_3^{\ast}=0.2$ 
corresponding to multimodal Gaussian mixture asymptotic
distribution $P(x,t\rightarrow \infty)$.
The supplementary view demonstrating the change of $P(x,t)$
is given in Fig.~\ref{fig:3dP}.

\begin{figure}[h!]
\includegraphics[scale=1.3]{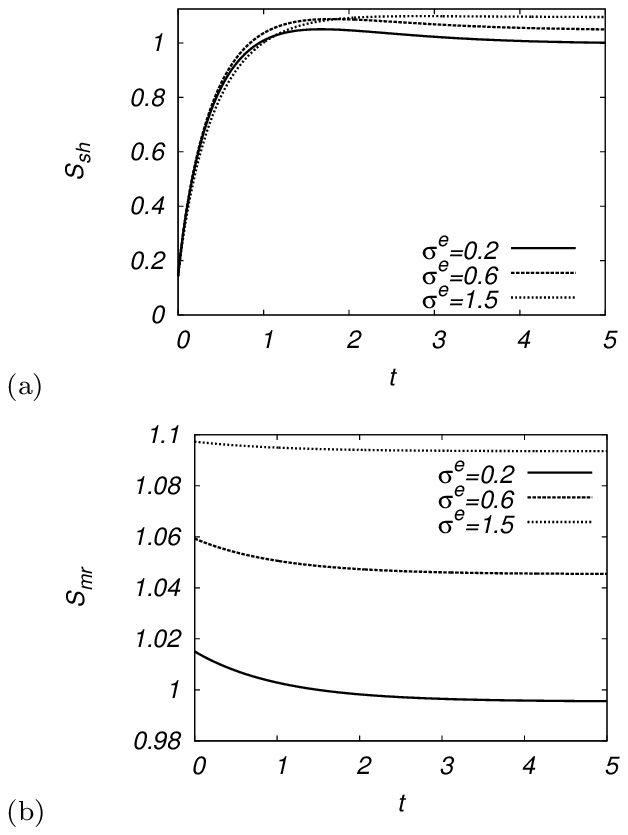} 
\caption{The non-equilibrium behavior of the Shannon entropy $S_{\rm sh}$ (part a) 
and Markovian entropy rate $S_{\rm mr}$ (part b). The complementary
dependencies to those presented  in Fig.~\ref{fig:3dP}) calculated
for perturbations - 3 distinct environments with the same 
mean $\mu^{\rm e}=1$, but different dispersion $\sigma^{\rm e} = 0.2, 0.6, 1.5$.
The calculation shows the non-monotonously growing entropy
(on the level of the clone) (see Eqs.~(\ref{S_sh},\ref{S_mr})). 
The alternative information-theoretic measure, the Markovian entropy rate shows
that transitions between attractors exhibit diversity decrease and saturation.}
\label{fig:entrop}
\end{figure}

The results of entropy variation as a function of time are depicted
in Fig.~\ref{fig:entrop}. They are obtained by substituting $p_i(t)$
into Eqs.~(\ref{S_sh}) and (\ref{S_mr}). 
For many systems the time dependence of entropy $S_{\rm sh}(t)$
is universal and consists in its initial increase, regarding not too many
details involved. The early diversification 
is followed by the subsequent speciation. As shown in part
b) of the figure, the relaxation does not affect significantly
the re-ordering extent (reflected by $S_{\rm mr}$) of the transitions
between states. Roughly speaking, the structure of states is more
persistent than their occupancy.

\section{Therapy as inverse problem}

Despite the exclusively theoretical nature of our work,
we outline its eventual contribution for therapy design.
When the therapeutic intervention is to consist in purposeful
manipulation with the statistics of the tumor microenvironment 
(represented here by the parameters $\mu^{\rm e},\sigma^{\rm e}$)
aimed to reduce heterogeneity ($\propto S_{\rm sh})$, the therapy
may be formally viewed as the entropy minimization problem
and solved by standard optimization techniques \cite{Floudas2009}.
To be more specific, the system of equations for $\sigma^{\rm e}(t)$
and $\mu^{\rm e}(t)$ can be written, that, in principle, represents 
steepest descent gradient dynamics towards their values 
in 'physiological' or 'desired' conditions, 
denoted by $\sigma^{\rm e}_{\rm des}$, $\mu^{\rm e}_{\rm des}$ 
and $s^{\rm e}_{\rm des}$
\begin{eqnarray}
\frac{{\rm d}
\ln\sigma^{\rm e}}{{\rm d}t} &=& - \frac{\partial d_{\rm ther}^2\,\,\,}{\partial \ln \sigma^{\rm e}}\,, 
\label{dsdt}
\\
\frac{{\rm d}
\mu^{\rm e}}{{\rm d}t} &=& - \frac{\partial d_{\rm ther}^2\,\,\,}{\partial \mu^{\rm e}}\,,
\nonumber 
\end{eqnarray}
which minimizes objective function in the form of 'therapeutic' weighted 
squared Euclidean distance 
\begin{eqnarray}
d_{\rm ther}^2&=& 
v_S \delta_S^2 +  
v_{\sigma} \delta_{\sigma}^2 + 
v_{\mu} \delta_{\mu}^2  
\label{eq:dther} 
\end{eqnarray}
consisting of the differences of desired and actual values of generalized coordinates  
\begin{eqnarray}
\delta_S &=& \ln S_{\rm sh,des} - \ln S_{\rm sh}\,,
\label{dsdt2}
\\
\delta_{\sigma} &=& \ln \sigma_{\rm des}^{\rm e} -  \ln \sigma^{\rm e}\,, 
\nonumber
\\
\delta_{\mu} &=& \mu_{\rm des}^{\rm e} -  \mu^{\rm e}\,, 
\nonumber
\end{eqnarray}
weighted by the respective constants $v_S>0$, $v_{\sigma}>0$, $v_{\mu}>0$. 
The term $\ln\sigma^{\rm e}$ is used instead of $\sigma^{\rm e}$ to 
keep the constraint $\sigma^{\rm e}\ge 0$ for any eventual solution.
Under some conditions, proper choice of desired entropy value
of $S_{\rm sh,des} \gtrsim 0$ can, in principle,
provide formally correct entropy decrease, prevented, however, by the effect
of $\delta_{\sigma}$, $\delta_{\mu}$, or by principal impossibility to 
decrease entropy beyond certain limits. The constant parameters $v_{\mu}$, 
$v_{\sigma}$ are used to bias optimization pressure from the values 
of $\mu^{\rm e}$, $\sigma^{\rm e}$ providing formally correct but, 
eventually, non-physiological solution ($\delta_S \sim 0$), towards their values 
$\mu_{\rm des}^{\rm e}$, $\sigma_{\rm des}^{\rm e}$ 
in standard physiological conditions. 

We emphasize that the above therapeutic considerations are implied
by applying the generalized distance-based concept as the crucial
aspect of our approach. In this context it is worth mentioning
that therapeutic model based on the construction of $d_{\rm ther}^2$
minimization pathway is inspired by the broad class of inverse
problems discussed in \cite{Tarantola2004}.
The above system of nonlinear equations (\ref{dsdt}) must be solved
simultaneously with equations for $\{ p_i(t)\}_{i=1}^N$ regarding
instant $S_{\rm sh}$.

\section{Conclusion}

Here outlined Markovian-based conceptualization links the uncertainty
of environment with intratumor heterogeneity, both expressed
in probabilistic terms. Evolutionary nature of carcinogenesis
\cite{Nowell1976} is respected, as the transition probabilities
correlate with statistical match between environment and
the attractors of the respective states, which corresponds
to the bet-hedging strategy \cite{Dejong2011}, evolved in
biological populations that face time-varying environment \cite{Muller2013}.

Recently, the evolutionary strategy called 'evolutionary trap' was
proposed \cite{Chen2015}. It consists of two steps, in which the first
stress 'channels' karyotypically divergent population into one
with a predominant drugable karyotypic feature, the second
stress targeting this feature \cite{Chen2015}.
Here presented approach follows the same aim, to lower
diversity of cancer cells population, in a more formal way.
Therapy is formulated as the optimization problem, representing
inverse modeling approach, which means evolving desired
phenotypic heterogeneity by purposeful manipulating with
the environment's statistics. Here, the 'desired' intratumor
heterogeneity corresponds to probability distribution represented
by the only peak, as narrow as possible. We believe that the
combination of the more formal approach, as proposed here,
with numerical simulations may provide interesting strategies,
going beyond usual intuition.

\vspace{10mm}

This work was supported by the (i) Scientific Grant Agency of the Ministry
of Education of Slovak Republic under the grant VEGA No. 1/0348/15, and
(ii) CELIM (316310) funded by 7FP EU (REGPOT).

\section*{Appendix: Regularization of Hellinger distance for the pair
of gaussian distributions} 

The purpose of this appendix is to review some elements of Hellinger distance calculus. 
We define the square of the Hellinger distance $d^2$ in terms of elementary probability
theory. If we denote the parametrization of the probability densities as $\phi(x,\theta_s)$, 
$s=1, 2$, then the squared Hellinger distance can be expressed as 
\begin{equation} 
d^2(\theta_1,\theta_2)  =  
1-  \int_{-\infty}^{\infty} \sqrt{\phi(x,\theta_1) \phi(x,\theta_2)} {\rm d}x\,.
\end{equation}
In the case of the pair of two Gaussian distributions 
$\phi_g(x,\mu_1,\sigma_1)$, $\phi_g(x,\mu_2,\sigma_2)$ constructed from the "template" 
\begin{equation}
\phi(x,\theta) \equiv \phi_g(x,\mu,\sigma)= \frac{1}{\sqrt{2\pi}\sigma} \exp\left[ - \frac{(x-\mu)^2}{2\sigma^2}\right]
\end{equation}
we obtained  
\begin{equation} 
\begin{split}
d^2(\mu_1,\sigma_1,\mu_2,\sigma_2) =
\ \ \ \ \ \ \ \ \ \ \ \ \ \ \ \ \ \ \ \
\ \ \ \ \ \ \ \ \ \ \ \ \ \ \ \ \ \ \ \ \ \ \ \ \\
1 - \sqrt{\frac{2 \sigma_1 \sigma_2}{ \sigma_1^2 + \sigma_2^2}} 
\exp\left( - \frac{ (\mu_1-\mu_2)^2 }{ 4 (\sigma_1^2 + \sigma_2^2) } 
\right)\,.
\end{split}
\end{equation} 
Since the limit $(\sigma_1, \sigma_2)\rightarrow (0,0)$ may create the
interpretation problems, the original form of $d^2(.)$ should be regularized. One possible way is to
use additive 
extra dispersion $\epsilon>0$ as follows 
\begin{equation} 
\begin{split}
d^2(\mu_1,\sigma_1,\mu_2, \sigma_2, \epsilon) =   
1 -  \sqrt{\frac{2 (\sigma_1+\epsilon) (\sigma_2+\epsilon)}{ 
(\sigma_1+\epsilon)^2 + (\sigma_2+\epsilon)^2}} \\
\times \exp\left( - \frac{ (\mu_1-\mu_2)^2 }{ 4 ((\sigma_1+\epsilon)^2 +(\sigma_2+\epsilon)^2) } 
\right)\,.
\end{split}
\end{equation} 
Thus, in the case when the 
original dispersions $\sigma_1, \sigma_2$ shrink to zero we have 
\begin{equation} 
d^2(\mu_1,0,\mu_2,0,\epsilon)  =  1 - 
\exp\left(-\frac{ (\mu_1-\mu_2)^2 }{ 8 \epsilon^2 }
\right)\,.
\end{equation} 
Then the Taylor expansion of the previously obtained function at $\mu_1 \sim \mu_2 $ yields 
\begin{equation} 
\begin{split}
d^2(\mu_1,0,\mu_2,0,\epsilon)  =
\ \ \ \ \ \ \ \ \ \ \ \ \ \ \ \ \ \ \ \
\ \ \ \ \ \ \ \ \ \ \ \ \ \ \ \ \ \ \ \ \ \ \ \ \\
\frac{(\mu_1-\mu_2)^2}{8 \epsilon^2} - 
\frac{(\mu_1-\mu_2)^4}{128 \epsilon^4}
+ {\cal O}((\mu_1-\mu_2)^6)\,
\end{split}
\end{equation}
with the leading term proportional to the dissimilarity measure analogous
to the one dimensional quadratic Euclidean squared distance $(\mu_1-\mu_2)^2$ 
between Cartesian coordinates $\mu_1$ and $\mu_2$ in 1d. Such demonstration
of the asymptotic consistency between generalized distance measure of the
probability distributions and classical analytical Euclidean distance in
1d supports the adequacy of $\epsilon>0$ regularization. 

The derivation highlights the interesting connection 
between traditional geometric and functional distance 
measures.  A further perspective in the analysis of
tumors consisting of several spatial compartments should
also be mentioned. In such case, the consequences of random 
switching could be readily quantified using Lukaszyk-Karmowski distance  
$\int {\rm d} \mu_1 \int {\rm d} \mu_2 \, \pi(\mu_1) 
\pi(\mu_2) \vert \mu_1  - \mu_2 \vert$  \cite{Lukaszyk2003}, 
which specifies geometric distance of the points with coordinates 
$\mu_1$ and $\mu_2$ known up to the respective probability
distributions $\pi(\mu_1)$, $\pi(\mu_2)$.

\end{document}